\documentclass{iopconfser}

\usepackage{acronym}
\usepackage{amsfonts}
\usepackage{amsmath}
\usepackage{amsthm} 
\usepackage{amssymb}
\usepackage{bbm} 
\usepackage{bm}
\usepackage{booktabs}
\usepackage{cases}
\usepackage{color}
\usepackage{comment}
\usepackage{dcolumn} 
\usepackage{enumerate}
\usepackage{enumitem}
\usepackage{epsfig}
\usepackage{etoolbox}
\usepackage{fancyref}
\usepackage{fancyhdr}
\usepackage[draft,author=,nomargin,inline]{fixme}
\usepackage[T1]{fontenc} 
\usepackage{footnote}
\usepackage{graphicx}
\usepackage{indentfirst}
\usepackage{latexsym}
\usepackage{lineno}
\usepackage{mathrsfs}
\usepackage{mciteplus}
\usepackage{multirow}
\usepackage{ragged2e}
\usepackage{rotating}
\usepackage{siunitx} % SI units 
\usepackage{scrextend}
\usepackage{soul}
\usepackage{subcaption} % subfigures with captions
\usepackage{verbatim}
\usepackage[colorlinks=true,linkcolor=blue,citecolor=blue,urlcolor=blue]{hyperref}
\usepackage{cleveref}

\makesavenoteenv{table} % Enable footnotes inside the table environment
\fxusetheme{color}
\FXRegisterAuthor{ml}{aml}{Michele}
\FXRegisterAuthor{ama}{aama}{Arnau}
\FXRegisterAuthor{cfs}{acfs}{Carlos}

\DeclareSIUnit{\solarmass}{\ensuremath{\mathit{M_{\odot}}}}
\DeclareSIUnit{\parsec}{pc}

\begin{document}

\title{Modified Black Hole Potentials and Their Korteweg-de Vries Integrals: Instabilities and Beyond}

\author{Michele Lenzi$^{1, 2}$, Arnau Montava Agudo$^{1, 3}$ and Carlos F. Sopuerta$^{1, 2}$}

\affil{$^1$ Institut de Ciències de l’Espai (ICE, CSIC), Campus UAB, Carrer de Can Magrans s/n, Cerdanyola del Vallès 08193, Spain}
\vspace{0.1cm}
\affil{$^2$ Institut d’Estudis Espacials de Catalunya (IEEC), Edifici RDIT, C/ Esteve Terradas, 1, desp. 212, Castelldefels 08860, Spain}
\vspace{0.1cm}
\affil{$^3$ Departament de Física, Universitat de les Illes Balears, IAC3–IEEC, Crta. Valldemossa km 7.5, E-07122 Palma, Spain}

\email{lenzi@ice.csic.es\,,\quad arnau.montava@uib.cat\,, \quad carlos.f.sopuerta@csic.es}

\begin{abstract}
Black Hole (BH) Quasi-Normal Modes (QNMs) and Greybody Factors (GBFs) are key signatures of BH dynamics that are crucial for testing fundamental physics via gravitational waves. Recent studies of the BH pseudospectrum have revealed instabilities in QNMs. Here, we introduce a new perspective using hidden symmetries in the BH dynamics, specifically the Korteweg-de Vries (KdV) integrals—an infinite series of conserved quantities. By analyzing modified BH potentials, we find strong evidence that KdV integrals are valuable indicators for detecting instabilities in QNMs and GBFs.
\end{abstract}

\section{Perturbed Black Holes: KdV Isospectral Symmetries and (in)Stability Studies}
The physical degrees of freedom of perturbed BHs can be described in terms of a set of master functions (one for each harmonic and parity) that satisfy a wave-like master equation (see, e.g.~\cite{Lenzi:2021wpc,Lenzi:2024tgk}):
\begin{equation}
\partial^2_t \Psi^{}_{\ell m} - \partial^2_{x} \Psi^{}_{\ell m} 
+ V^{}_\ell(r)\, \Psi^{}_{\ell m} = 0\,,
\label{master-equation}
\end{equation}
where $\Psi_{\ell m}$ is the master function, with harmonic numbers $(\ell, m)$, $V_\ell$ is the potential, and $x$ the tortoise coordinate. These master equations possess Darboux and KdV symmetries~\cite{Lenzi:2021njy,Lenzi:2022wjv,Lenzi:2023inn,Jaramillo:2024qjz}. In particular, if we deform the time-independent version of Eq.~\eqref{master-equation} along the flow of the KdV equation (see, e.g., \cite{faddeev1976:jsm,Lenzi:2025kqs})
\begin{equation}
V^{}_{,\tau} = 6 V V^{}_{,x} - V^{}_{,xxx} \,,
\end{equation}
where $\tau$ is the deformation parameter,  the spectrum is preserved, that is, the BH QNMs and GBFs are preserved. There is a connection between the conserved quantities of these symmetries, the KdV integrals $K_{n}$, and the moments of a functional of the BH transmission coefficient $T(k)$, via the \emph{trace identities}~\cite{Lenzi:2022wjv,Lenzi:2023inn}:
\begin{equation}
(-1)^{n+1} 2^{-2n}\pi K^{}_{2n+1} = \int_{-\infty}^{\infty} dk \; k^{2n} \ln T(k) \,.
\label{Eq:Trace-indentities}
\end{equation}

\begin{figure}[h!]
\centering
\begin{subfigure}{0.35\textwidth}
    \centering
    \includegraphics[width=0.8\textwidth]{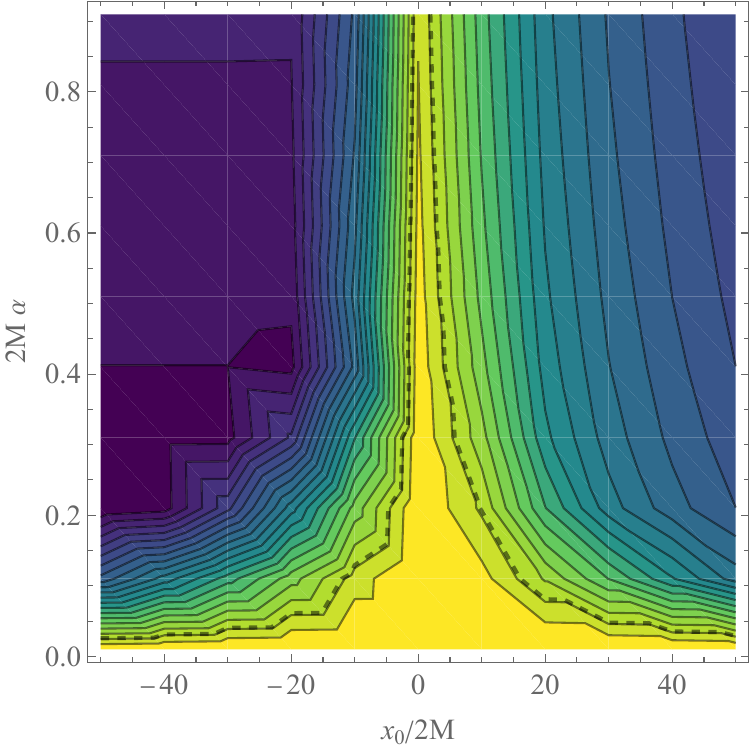}
    \includegraphics[width=0.16\textwidth]{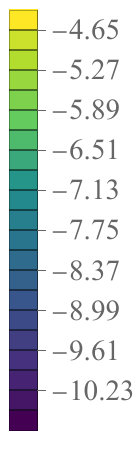}
    %\caption{} % optional mini-caption
%\label{Fig:KdVbump_x0-alpha}
\end{subfigure} \qquad \qquad
%\hfill
\begin{subfigure}{0.38\textwidth}
    \centering
    \includegraphics[width=\textwidth]{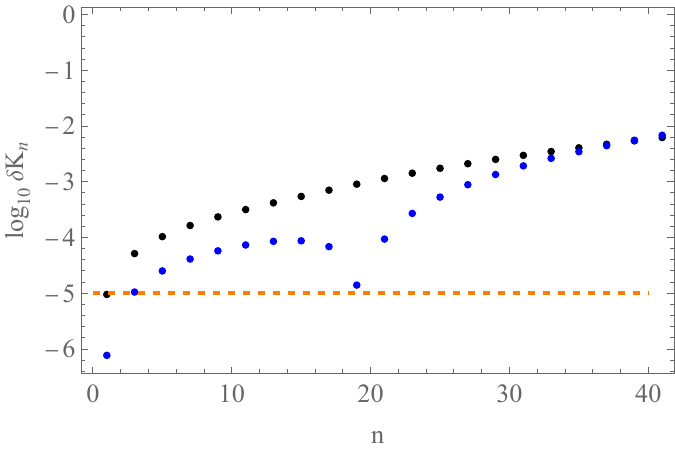}
    %\caption{} % optional mini-caption
%\label{Fig:KdV_EFT_n}
\end{subfigure}
\caption{(Left) Relative error $\log_{10}\delta \mathcal{K}_{5}$ for $\epsilon = 10^{-5}$ and varying $x_0$ and $\alpha$. The dashed line corresponds to the threshold $\delta \mathcal{K}_{n}=\epsilon$. 
(Right) Relative error $\delta \mathcal{K}_{2n-1}$ for $n=1,...,30$ for the odd (blue dots) and even (black dots) potentials. The dashed line corresponds to the threshold $\epsilon=10^{-5}$ (orange dashed line). 
}
\label{Fig:KdV_combined}
\end{figure}

Here, we advocate~\cite{Lenzi:2025kqs} for the use of KdV integrals as indicators of BH spectral properties, with focus on isospectrality between even- and odd-parity perturbations and QNM/GBF instabilities \cite{jaramillo2021:prx}. In Fig.~\ref{Fig:KdV_combined} we present the stability properties of the KdV integrals~\cite{Lenzi:2025kqs} for some of the modified BH  potentials studied in the literature, for both odd and even parities, which have the following structure:
\begin{equation}
V^{\mathrm{odd/even}} \;=\; V^{\text{RW/Z}} \;+\; \epsilon\,\delta V^{\text{odd/even}} \,,
\end{equation}
where $\epsilon\ll 1$, and the superscripts 'RW' and 'Z' denote the Regge-Wheeler~\cite{Regge:1957td} and Zerilli~\cite{Zerilli:1970la} BH potentials.
In the left panel of Fig.~\ref{Fig:KdV_combined}, motivated by the form of some astrophysical environmental effects,  the potential modification is a P\"oschl-Teller bump with width $\alpha$ and centered around $x_0$: 

\begin{equation}
\delta V \;=\; r^{-2}_s \operatorname{sech}^2\!\big[\alpha\,(x-x_0)\big]
\;=\; r^{-2}_s \,\cosh^{-2}\!\big[\alpha\,(x-x_0)\big] \,,
\label{Eq:Poschl-Teller-bump}
\end{equation}

where $r_s$ denotes the Schwarzschild radius.
In the right panel of Fig.~\ref{Fig:KdV_combined}, we plot potential modifications coming from an Effective-Field Theory (EFT)~\cite{silva2024:prd}:
\begin{equation}
\delta V^{\text{odd}} \;=\; \frac{1}{r^{2}}\sum_{i=1}^{7} v^{\text{odd}}_{i}\!\left(\frac{M}{r}\right)^{i},
\qquad
\delta V^{\text{even}} \;=\; \frac{4}{\bigl(\lambda r\bigr)^{2}}\sum_{i=1}^{10} v^{\text{even}}_{i}\!\left(\frac{M}{r}\right)^{i} .
\end{equation}
For the two potential modifications we study the relative error in the KdV integrals. 

The relative error, $\delta K_n$, and the corresponding stability criterion are:
\begin{equation}
\delta K^{}_n \;=\;
\left|\,\frac{K_n^{\epsilon} - K_n^{\mathrm{RW}}}{K_n^{\mathrm{RW}}}\,\right| \,,
\qquad \qquad
\delta K^{}_n \;\lesssim\; \epsilon \,,
\end{equation}
where here $K_{n}^{RW /\epsilon}$ is the n-th KdV integral of the Regge-Wheeler or of the modified potential, respectively.

The general trend emerging from this analysis (see~\cite{Lenzi:2025kqs} for the detailed development) shows that high-order KdV integrals tend to be more sensitive to small scale perturbations while the lower-order ones are destabilized by wide perturbations.  
Moreover, the KdV integrals appear to be an important and simple indicator of the loss of isospectrality between the odd- and even-parity sectors in the EFT case.

\section{Greybody Factor Stability}
\begin{figure}[h!]
\centering
\begin{subfigure}{0.35\textwidth}
    \centering
    \includegraphics[width=\textwidth]{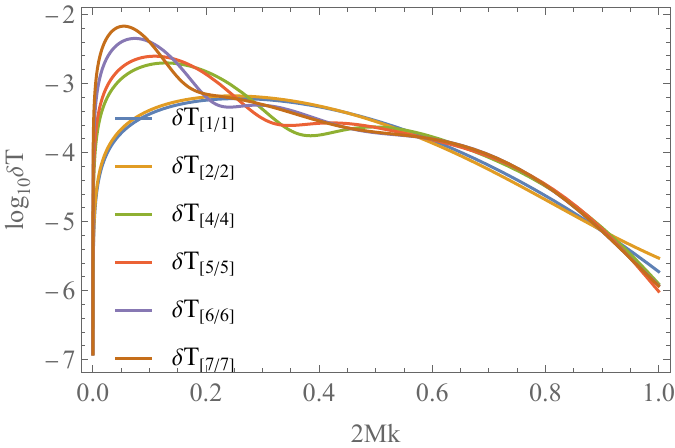}
%\caption{} % optional mini-caption
%\label{Fig:GF-diag-subdiag}
\end{subfigure}
%\hfill 
\qquad \qquad
\begin{subfigure}{0.36\textwidth}
    \centering
    \includegraphics[width=\textwidth]{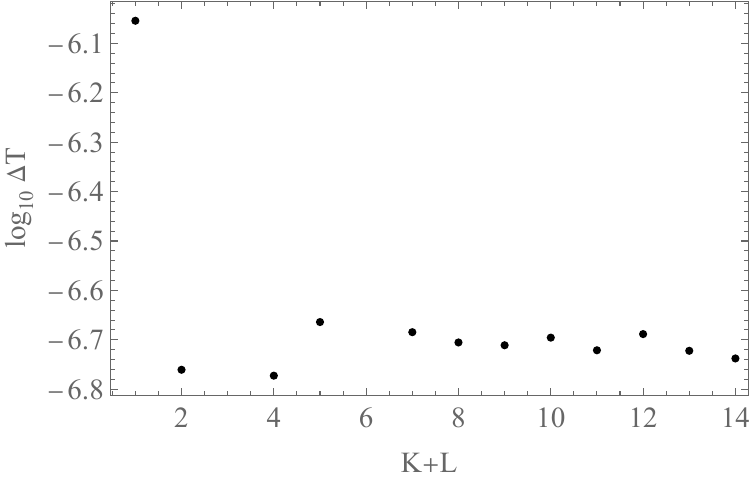}
%\label{Fig:GF-error-integ}
\end{subfigure}
\caption{(Left) Relative error for the P\"oschl–Teller bump corrections for $r^{}_s\,\alpha=1/50$, $x_0=0$ and $\epsilon=10^{-5}$ for diagonal Pad\'e approximants. 
(Right) Integrated error for the same parameters in the left plot. On the $x$-axis we have $K+L$, tracking the order of the Pad\'e approximants.}
\label{Fig:GF_combined}
\end{figure}

The KdV integrals can be seen as the first integrals of an associated Hamiltonian system~\cite{faddeev1976:jsm}, and in this way are obtained by integration in configuration space. In contrast, as the trace identities show [see Eq.~\eqref{Eq:Trace-indentities}], they can also be seen as the moments of a
frequency-dependent distribution function that is logarithmic in the BH transmission probability, one of the GBFs, and then they can be obtained as integrals in the frequency domain.
This fact helps us to understand the interplay between the stability properties of the KdV integrals and the stability of the GBFs, in particular the transmission coefficient $T(k)$. We have studied the stability of the GBFs, evaluated only in terms of the KdV integrals with Pad\'e approximants~\cite{Lenzi:2022wjv,Lenzi:2023inn}, by using the relative and integrated errors $\delta T_{[N/M]}$ and $\Delta T_{[N/M]}$ respectively:
\begin{equation}
\delta T_{[N/M]} \;=\;
\frac{\bigl|\,T^{\text{bump}}_{[N/M]} - T^{\mathrm{RW}}_{[N/M]}\,\bigr|}{T^{\mathrm{RW}}_{[N/M]}} \,,
\qquad \qquad
\Delta T_{[N/M]} \;=\;
\frac{\displaystyle \int_{0}^{\infty}\! dk \;\bigl|\,T^{\text{bump}}_{[N/M]} - T^{\mathrm{RW}}_{[N/M]}\,\bigr|}
{\displaystyle \int_{0}^{\infty}\! dk \; T^{\mathrm{RW}}_{[N/M]}} \,,
\end{equation}
where the brackets $[N/M]$ make reference to the order of the Padé approximant considered, which are at the heart of the solution of the moment problem defined by the trace identities of Eq.~\eqref{Eq:Trace-indentities} (see Refs.~\cite{Lenzi:2022wjv,Lenzi:2023inn}).
In Fig.~\ref{Fig:GF_combined} we plot both the relative (left panel) and integrated errors (right panel), for the Pöschl-Teller bump correction introduced in Eq.~\eqref{Eq:Poschl-Teller-bump}.

What we observe from this study is that the stability or instability of low- and high-order KdV integrals correctly accounts for the stability or instability of the logarithm of the GBFs both at low and high frequencies. Furthermore, the instability of the KdV integrals only affects the GBFs locally, while the integrated error remains well below the instability zone.

\section{Conclusions}
We investigated the possible relation between the integrable KdV structures appearing in the description of BH perturbations and some spectral properties of the system, namely GBFs and QNMs.
Our analysis, presented in detail in~\cite{Lenzi:2025kqs}, highlights several key insights into the relation between the BH potentials, the KdV integrals, and the GBF. We have found that the first few KdV integrals are highly sensitive to long-range (infrared) modifications of the BH potential, whereas the strongest instabilities arise in the higher-order KdV integrals, which are primarily driven by local/short-range features of the potential.

Considering the point of view in which the KdV integrals are the moments of a frequency-domain distribution clarifies the pattern: the instability of the first few moments (and KdV integrals) reflects the infrared sensitivity characteristic of low-order moments, whereas higher-order moments respond to high-frequency (ultraviolet) modifications. From this moment problem viewpoint~\cite{Lenzi:2022wjv,Lenzi:2023inn}, the high-frequency instability of the reflectivity GBF (as shown in~\cite{Oshita:2024fzf}), $R(k)$ -with $R(k)+T(k)=1$ at the probability level- is naturally explained by, and linked to, the instability of the higher-order KdV integrals (as analyzed in detail in~\cite{Lenzi:2025kqs}).

Finally, we observe a qualitative connection between the hierarchy of KdV integrals and the BH QNM instabilities, suggesting a structural bridge between scattering data and the spectrum of the BH perturbations. This link motivates using the KdV hierarchy of conserved quantities and the associated moment problem perspective as solid diagnostics to study how potential deformations propagate to both GBFs factors and the QNM behavior.

%%%%% ACKNOWLEDGMENTS
\section*{Acknowledgments}
ML and CFS have been supported by contract PID2019-106515GB-I00/AEI/10.13039/501100011033 (Spanish Ministry of Science and Innovation, MCIN) and by the program Unidad de Excelencia María de Maeztu CEX2020-001058-M (MCIN). ML has also been supported by a Juan de la Cierva contract FJC2021-047289-I funded by program MCIN/AEI/10.13039/501100011033 (MCIN) and by NextGenerationEU/PRTR (European Union). AMA was supported by the JAE-ICU program (2023-ICE-07) linked to the María de Maeztu contract CEX2020-001058-M and also by the Programme RI24- researcher in gravitational-wave astronomy (Research and Innovation Program linked to Component 23, Investment 1, within the framework of the Recovery, Transformation and Resilience Plan, funded by the European Union– Next Generation EU, corresponding to the years 2024 and 2025). AMA was also supported by the Spanish Agencia Estatal de Investigación grants PID2022-138626NB-I00, RED2024-153978-E, RED2024-153735-E, funded by MICIU/AEI/10.13039/501100011033 and the ERDF/EU; and the Comunitat Autònoma de les Illes Balears through the Conselleria d'Educació i Universitats with funds from the European Union - NextGenerationEU/PRTR-C17.I1 (SINCO2022/6719) and from the European Union - European Regional Development Fund (ERDF) (SINCO2022/18146).

%%%%% BIBLIOGRAPHY
\bibliographystyle{iopart-num}
\bibliography{iopart-num}

\end{document}